\documentclass[fleqn,usenatbib]{mnras}

\usepackage{newtxtext,newtxmath}

\usepackage[T1]{fontenc}

\DeclareRobustCommand{\VAN}[3]{#2}
\let\VANthebibliography\thebibliography
\def\thebibliography{\DeclareRobustCommand{\VAN}[3]{##3}\VANthebibliography}

\usepackage{graphicx}	
\usepackage{amsmath}	

\def\P     {P$_{1400}$ }

\def\le    {L$_{\rm Edd}$ }

\def \arcmin      {\text{$^\prime$}}
\def \arcsec      {\text{$^{\prime\prime}$}}

\def \mujybeam    {$\muup$Jy\,beam$^{-1}$}

\title[RAD@home discovery of a bow-and-arrow radio galaxy (BAARG).]
{RAD@home discovery of a bow-and-arrow radio galaxy tracing a $\sim$560 kpc bow-shock structure in a multi-halo environment}

\author[Ananda Hota et al.]{Ananda Hota,$^{1,2,3}$\thanks{E-mail:hotaananda@gmail.com}
Pratik Dabhade,$^{4,3}$ Shubhrangshu Ghosh,$^{5,6,3}$ Pranim Limbo,$^{3}$  C. Konar,$^{7,3}$ 
\newauthor 
Sagar Sethi,$^{8,3}$ Souvik Manik,$^{9}$ Aditya Sahasranshu,$^{3}$ Sabyasachi Pal,$^{9,3}$ Mitali Damle,$^{10,11,3}$ 
\newauthor 
Sravani Vaddi,$^{3}$ Arundhati Purohit$^{3}$ 
\\
$^{1}$ \#eAstroLab, UM-DAE Centre for Excellence in Basic Sciences, University of Mumbai, Santacruz-East, Mumbai-400098, India\\ 
$^{2}$ Centre for Excellence in Theoretical and Computational Science, University of Mumbai, Santacruz-East, Mumbai-400098, India\\
$^{3}$ RAD@home Astronomy Collaboratory, Kharghar, Navi Mumbai, PIN 410210, India\\
$^{4}$ Astrophysics Division, National Centre for Nuclear Research, Pasteura 7, 02-093 Warsaw, Poland \\
$^{5}$ Center for Astrophysics, Gravitation and Cosmology (CAGC), Shri Ramasamy Memorial (SRM) University Sikkim, Gangtok, 737102, India\\
$^{6}$ Department of Physics, Shri Ramasamy Memorial (SRM) University Sikkim, 5th Mile Tadong, Gangtok, 737102, India\\
$^{7}$ Amity Institute of Applied Sciences, Amity University Uttar Pradesh, Sector-125, Noida-201303, India\\
$^{8}$ Space Radio‐Diagnostics Research Centre, University of Warmia and Mazury, R. Prawochenskiego 9, 10‐719 Olsztyn, Poland\\
$^{9}$ Midnapore City College, Kuturia, Bhadutala, Paschim Medinipur, West Bengal-721129, India\\ 
$^{10}$ New York University Abu Dhabi, Department of Physics, PO Box 129188, Abu Dhabi, UAE\\
$^{11}$ Center for Astrophysics and Space Science (CASS), New York University Abu Dhabi, PO Box 129188, Abu Dhabi, UAE
}

\pubyear{2026}

\begin{document}
\label{firstpage}
\pagerange{\pageref{firstpage}--\pageref{lastpage}}
\maketitle
\begin{abstract}

We report the RAD@home citizen science discovery of a unique bow-and-arrow-shaped radio galaxy (BAARG; RAD J104501.6$+$352852, $z=0.159$) identified in LoTSS~DR2. The source exhibits striking asymmetry: on the western side, a narrow jet feeds a sector-shaped emission region at $\sim$115\,kpc, extending backward to form a $\sim$560\,kpc arc-like structure; on the eastern side, the jet develops an S-shaped distortion extending to $\sim$250\,kpc, followed by a faint, offset tail reaching $\sim$600\,kpc.  Our analysis shows that the host resides in a dynamically complex, multi-halo environment with nearby cluster-scale systems at similar redshifts. The observed morphology is consistent with interaction between the radio plasma and the surrounding medium, influenced by large-scale environmental gradients and bulk motions. The western structure is consistent with compression of radio plasma near a bow-shock-like feature, possibly linked to supersonic motion of the infalling host galaxy and its circumgalactic medium. This possibly represents one of the first instances in which morphology and environment together suggest signatures of infall- or shock-related processes; surveys such as LoTSS~DR3 may reveal similar systems, offering new insights into the interplay between radio galaxies and their large-scale environments.

\end{abstract}



\section{Introduction}\label{sec:1_intro}
Radio galaxies (RGs) typically exhibit synchrotron-emitting, non-thermal radio lobes that are largely uncorrelated with the thermal stellar structures seen in optical images, apart from the host galaxy itself. The advent of high-resolution X-ray observatories such as \textit{Chandra} and \textit{XMM-Newton} has established that radio lobes interact strongly with the surrounding intracluster medium, often inflating cavities co-spatial with the radio emission \citep[e.g.][]{HydraA-cavity-McNamara2000, cavity-Birzan2004, McNamara07}.
Low-frequency observations with the GMRT and LOFAR \citep[e.g.,][]{Sullivan2011,Timmerman2022}, complemented by high-frequency data from the VLA, have shown that many X-ray cavities are filled with steep-spectrum, aged synchrotron plasma, which are relics of past AGN feedback episodes. X-ray and radio observations have established that expanding radio lobes, primarily from Fanaroff-Riley type II \citep[FR\,II;][]{FR74} RGs, shock-heat the surrounding intracluster medium (ICM); in these systems, the shocked gas surrounding the lobes exhibits higher temperatures than the ambient, pre-shocked gas. In specific cases, such as the nearby radio galaxy Cen A, X-ray imaging directly resolves the shocked gas immediately ahead of the non-thermal radio lobe \citep{CenA-Croston-2009}. The X-ray emission from this shock shell follows a clear power-law spectrum, confirming its non-thermal origin.

With the advent of high-sensitivity, high-dynamic-range interferometric radio observations, a range of previously unrecognised features have been identified in RGs and their surrounding environments.
Beyond standard components such as core, jet, hotspot, and lobes, these new features include filaments \citep{dancing-ghosts-Velovic2023}, giant kinks \citep{2022A&A...668A..64D}, Collimated Synchrotron Threads \citep[CSTs;][]{CST-Ramatsoku2020, Condon21IC4296, RAD-ISRA-Hota2025}, radio rings, and potentially Odd Radio Circles \citep[ORCs;][]{2022Norris, RAD-ORC-Hota2025}, large-scale box-shaped bubbles on different scales due to multiple jet episodes \citep{nest-oldest-relic-Brienza2021Nature}. Further examples include off-axis jet diversions \citep[e.g., 3C 321;][]{jet-galaxy-3C321-Evans2008} and mushroom-shaped lobes \citep[e.g., RAD12;][]{Hota2022RAD12} resulting from jet–galaxy interactions. Similarly, many peculiar-morphology radio sources that have been identified recently from various large-scale surveys may also be hiding new physical processes in and around radio galaxies \citep[e.g. ][]{complex-sources-LoTSS-Horton2025, strange-sources-Sasmal2025, ORC-Gupta-2025PASA, RAD-ISRA-Hota2025}.  Consequently, radio morphologies are no longer strictly defined by linear jets, precessional changes \citep[S-, Z-, or X-shaped sources;][]{GK2003,Bera2020,Sethi2024}, or ram-pressure-induced curvature, such as head-tail and wide-angle tail galaxies \citep[e.g.,][]{OdeaWAT2023}. Instead, we may be observing pre-existing magnetic and hydrodynamic structures and dynamical processes in the circum-galactic medium, extending up to hundreds of kiloparsecs from the host, made visible by the plasma supplied by AGN-driven jets.

The identification and interpretation of such features rely on combining radio data across resolutions and frequencies with optical imaging and redshift information, enabling reliable association with host galaxies, determination of their physical properties, and characterisation of their large-scale environments \citep[e.g.,][]{Sankhyayan2024,Mahato2026}.

The radio-optical multi-frequency association is one of the fundamental skills taught to students and participants in citizen-science initiatives such as RAD@home \citep{Hota2014,Hota16} and Radio Galaxy Zoo \citep{RGZ-Banfield_2015}. Trained citizen scientists can identify rare sources with faint features that automated machine-learning algorithms often overlook. For example, the highest-redshift and most powerful twin-ring ORC discovered by RAD@home \citep{RAD-ORC-Hota2025} was previously misclassified as a giant radio galaxy (GRG\footnote{Defined as RGs with overall sizes $\geq$700~kpc \citep{GRGreview}}) in automated studies \citep{RAD-ORC-GRG-Mostert2024}, highlighting the challenges in identifying complex and rare radio morphologies. When source structures are complex or extremely faint, automated algorithms often fail due to a lack of specific training sets. In contrast, citizen scientists, unbiased by algorithmic constraints, can synthesise morphological details, optical colours, and redshifts to develop an intuitive understanding of the dynamical evolution of a galaxy, leading to the discovery of both novel structures and their associated physical phenomena. 

In this paper, we present the discovery of a uniquely structured radio galaxy, the ``Bow-and-Arrow-shaped Radio Galaxy'' (BAARG), identified through the RAD@home citizen-science collaboratory. This source provides a rare look at the interaction between a galaxy and its environment. 
Its morphology is characterised by a large-scale bow shock, revealing a characteristic arc-shaped structure, likely forming at the periphery of the circumgalactic medium (CGM) as the host galaxy moves through the ICM. By analysing these features, we demonstrate how jet-supplied plasma can act as a ``tracer'', illuminating the hidden hydrodynamic boundaries and shock fronts created by a galaxy's motion through a dense cluster environment.

In this paper, we adopt the flat $\Lambda$CDM cosmological model based on the Planck results \citep{Planck}, with parameters $H_0$ = 67.4 km s$^{-1}$ Mpc$^{-1}$ (Hubble-Lema\^itre constant), $\Omega_m$ = 0.315, and $\Omega_{\Lambda}$ = 0.685 . The following radio-spectral index convention is being followed: \( S_{\nu} \propto \nu^{\alpha} \), where \( S_{\nu} \) is the flux density at frequency \( \nu \) and \( \alpha \) is the spectral index. The coordinates in all images are in the J2000 system. 

\vspace{-0.5cm}
\section{Citizen science discovery process}\label{sec:discovery}

RAD@home\footnote{\url{https://www.radathomeindia.org}} is a citizen-science research collaboratory founded in 2013 \citep{Hota2014} as a zero-funding, zero-infrastructure, inter-university network. It links professional astronomers with a nationwide pool of trained citizen scientists across India \citep[for more details, see][]{Hota16}. Participants are selected for long-term training, ranging from one week to several months, following their participation in one-day workshops hosted by various research and educational institutes. These volunteers are trained in the multi-wavelength interpretation of galaxies using the RAD@home RGB-maker, a web-based tool that generates red-green-blue-contour (RGB-C) images by integrating UV, optical, IR, and radio data from various all-sky surveys \citep{2023IAUS..375...40K}. After this initial training, participants discuss their findings from radio survey data during online e-classes held on weekends. Among the diverse rare radio sources identified by the collaboratory \citep{RAD-ISRA-Hota2025}, two recent discoveries are particularly significant: the jet–galaxy interaction in RAD12 \citep{Hota2022RAD12} and the highest-redshift and most luminous ORC RAD J131346.9$+$500320 \citep{RAD-ORC-Hota2025}.

\vspace{-0.5cm}
\section{Results}
\begin{figure*}
    \centering
     \includegraphics[scale=0.65]{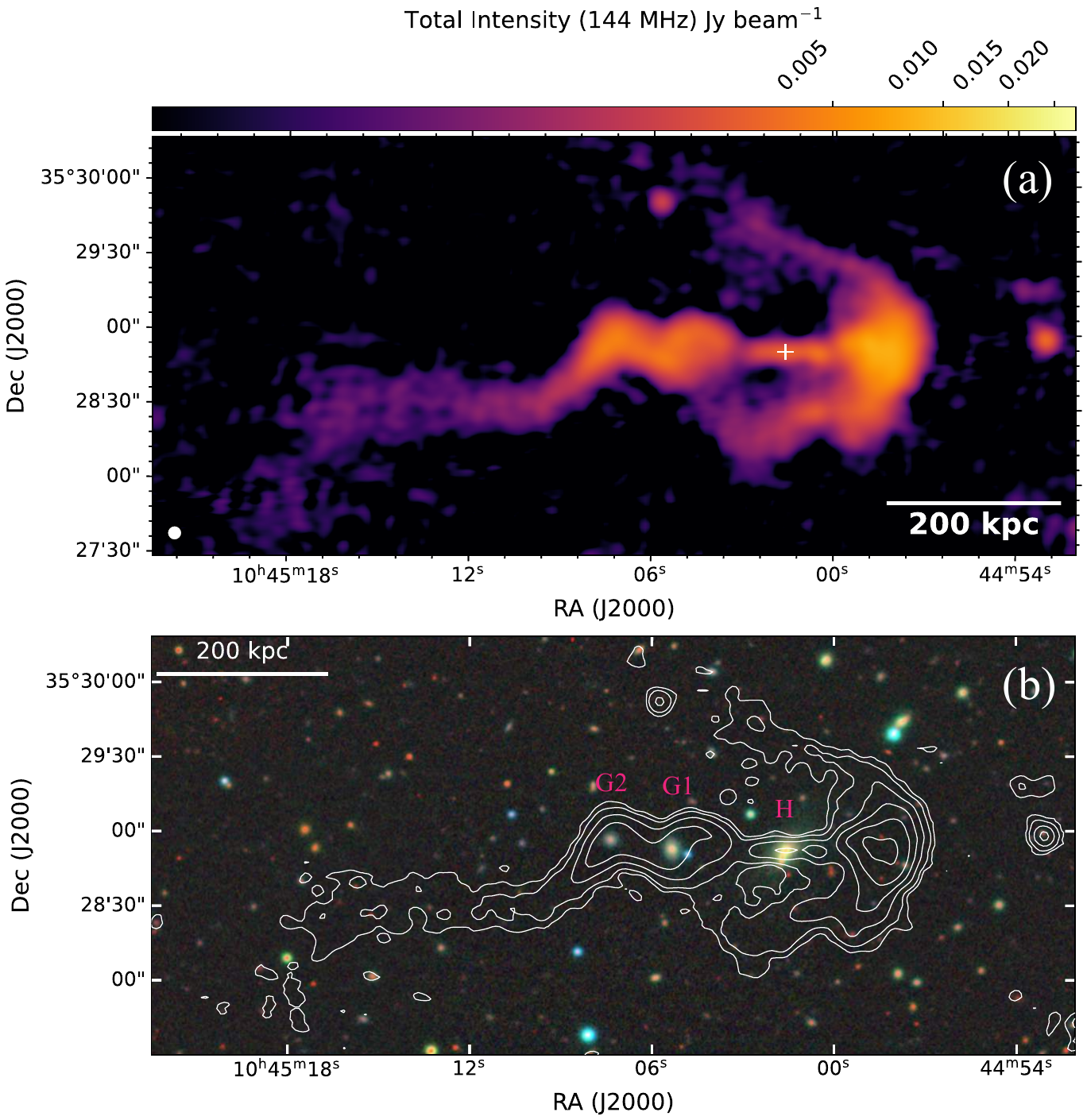}
     \caption{Upper: LoTSS 144~MHz radio image with 6\arcsec angular resolution of BAARG, where only emission above $3\sigma$ is shown, and the $\sigma$ (rms noise) is $\sim$\,71 \mujybeam. The white cross shows the location of the object's host galaxy. Lower: LoTSS 144~MHz radio contours for emission above 3$\sigma$ ($10^{-3} \times$ [0.21,0.49,1.35,2.78,4.48,7.73,10.51] Jy~beam$^{-1}$) at $6^{\prime\prime}$ angular resolution in white, overlaid on the BASS optical image. Zoomed-in higher resolution optical images of galaxies marked as H, G1, and G2 are presented in Fig.~\ref{fig:O_HSCLA}.}
    \label{fig:Bow-and-Arrow-LoTSS-on-BASS}
\end{figure*}

\subsection{BAARG: Bow-and-Arrow-shaped radio galaxy } \label{sec:Bow}
A unique radio source with a bow-and-arrow-like structure was first identified by a citizen scientist during an online weekend e-class on 24$^{\rm th}$ May 2025. The striking bow-and-arrow-like morphology is visible only in the LoTSS~DR2 144~MHz data \citep{lotssshimwell}, and it appears as double lobed in coarser resolution surveys like NRAO VLA Sky Survey map \citep[NVSS:][]{nvss} and TIFR GMRT Sky Survey \citep[TGSS:][]{tgss_intema}. The source is only partially detected in the FIRST survey \citep{Helfandfirst}. The source has previously been catalogued as a GRG \citep{BAARG-GRG-LOFAR-Oei2023}; however, its detailed morphology is presented here for the first time using LoTSS.

The LoTSS image (Fig.~\ref{fig:Bow-and-Arrow-LoTSS-on-BASS}a) reveals fine structural details that do not conform to standard RG morphologies. The total flux density and radio power at 144 MHz are  $\sim\,$320~mJy and $2.2\times 10^{25} \rm W~Hz^{-1}$, respectively. As shown in Fig.~\ref{fig:Bow-and-Arrow-LoTSS-on-BASS}a, the source is neither edge-brightened (FR\,II) nor edge-darkened (FR\,I). On small scales, the radio jets extend approximately east–west and exhibit comparable brightness on either side of the host galaxy. At a distance of $\sim$\,41\arcsec\ or 115 kpc, the jet flares into a sector-shaped conical region that is brighter than the jet itself. The curved edge of this sector extends further to form a faint but large bow-shaped structure. The north-western half of the bow feature extends nearly 100\arcsec\ ($\sim$280 kpc), implying a total projected size of $\sim$560 kpc for the full shell-like or bow-shock-like structure. The south-western portion appears somewhat disturbed, although a comparable linear extent to the north-western side is evident. Notably, this bow-like structure extends well beyond the host galaxy, unlike the backflow or bridge emission typically seen in FR\,II RGs, suggesting a different dynamical process for the formation of the bow-shaped structure. 

On the eastern side, the jet develops an S-shaped distortion, extending up to $\sim$\,250 kpc (from the host galaxy to the end of the S-shaped region). Interestingly, regions of the S-structure are brighter than the initial jet/plume emerging from the host galaxy. The radio plume then shifts southwards (undetected in NVSS \& TGSS, see Fig.~\ref{fig:rgbm-specin}), from the initial jet direction and extends up to 215\arcsec\ ($\sim$\,600 kpc) from the host galaxy. As seen in the LoTSS map (Fig.~\ref{fig:Bow-and-Arrow-LoTSS-on-BASS}a), the western side exhibits features resembling those of FR\,II RGs, while the eastern side shows clear similarity to edge-darkened FR\,I morphologies. Consequently, the overall structure is reminiscent of Hybrid Morphology Radio Sources \citep[HyMoRS;][]{HyMoRS-Gopal2000, kapinska17, HyMoRS-Manik2026}. However, the presence of a large-scale bow-shaped feature extending beyond the lobe distinguishes this source from typical HyMoRS. We refer to this source as BAARG.

The Fig.~\ref{fig:Bow-and-Arrow-LoTSS-on-BASS}b presents LoTSS radio contours overlaid on the colour RGB image from the Beijing–Arizona Sky Survey \citep[BASS;][]{BASS}. The host galaxy (SDSS J104501.61+352852.2; u=20.52$\pm$0.14, r=17.32$\pm$0.01, $z_{\rm spec}=0.15939\pm0.00004$) is clearly a red ($u-r=3.2$) elliptical. On close inspection using the deeper and higher resolution g-band image from the Hyper Suprime-Cam Legacy Archive \citep[HSCLA;][]{HSCLA-Tanaka2021}, it can be seen that there is another galaxy (marked as N) overlapping with the host galaxy (see panel `H' of Fig.~\ref{fig:O_HSCLA} in the Appendix.~\ref{sec:app-ims}). This nearby galaxy (N) is an edge-on lenticular or disc galaxy (SDSS J104501.72$+$352849.1, $z_{\rm phot}=0.181\pm0.018$). Given the substantial difference in redshift between the lenticular galaxy and the BAARG host, and the undisturbed morphology of the lenticular disc, it is highly unlikely that the BAARG host and this galaxy are interacting. On the western side, the radio peak connected to the bow structure shows no co-spatial optical galaxy, ruling out the possibility that the bow represents an independent WAT RG. This western peak, which forms part of the sector-shaped conical region, differs markedly from the lobes of typical FRII sources with hotspot and backflow features. 

The eastern plume, on the other hand, shows two optical galaxies (see panels G1 \& G2 in Fig.~\ref{fig:O_HSCLA}) coinciding with the radio peaks of the S-shaped structure (also see Fig.~\ref{fig:Bow-and-Arrow-LoTSS-on-BASS}b). G1 (SDSS~J104505.35$+$352852.7) is an inclined spiral with $z_{\rm phot}=0.152\pm0.020$ (similar to BAARG), which appears to be interacting first with the jet. Owing to the uncertainties in photometric redshift and the lack of high-resolution, deep radio imaging to robustly characterise the interaction, this scenario remains uncertain, although it cannot be ruled out. G2 (SDSS~J104507.37$+$352856.4), with $z_{\rm phot}=0.196\pm0.046$, is co-spatial with another radio peak of the S-shaped structure. Beyond G2, the radio plume deviates southward while remaining broadly aligned with the initial jet direction, extending up to $\sim$600 kpc from the host galaxy. This part of the S-shaped structure appears morphologically peculiar and may suggest another possible jet-galaxy interaction; however, the significant redshift difference between the BAARG host and G2 argues against a physical association. The combined extent of the eastern tail and the western sector region yields a total projected size of $\sim$715 kpc, placing the source in the GRG class.

The western structure has an integrated flux density of 195~mJy, compared to 125~mJy for the eastern side. The western side is more compact and brighter, while the eastern side is more extended and fainter, despite comparable jet brightness near the host. This asymmetry suggests a significant influence of the large-scale environment, which we discuss in the following section.

\vspace{-0.65cm}
\subsection{Large-Scale Environment of BAARG}
\begin{figure}
\includegraphics[scale=0.428]{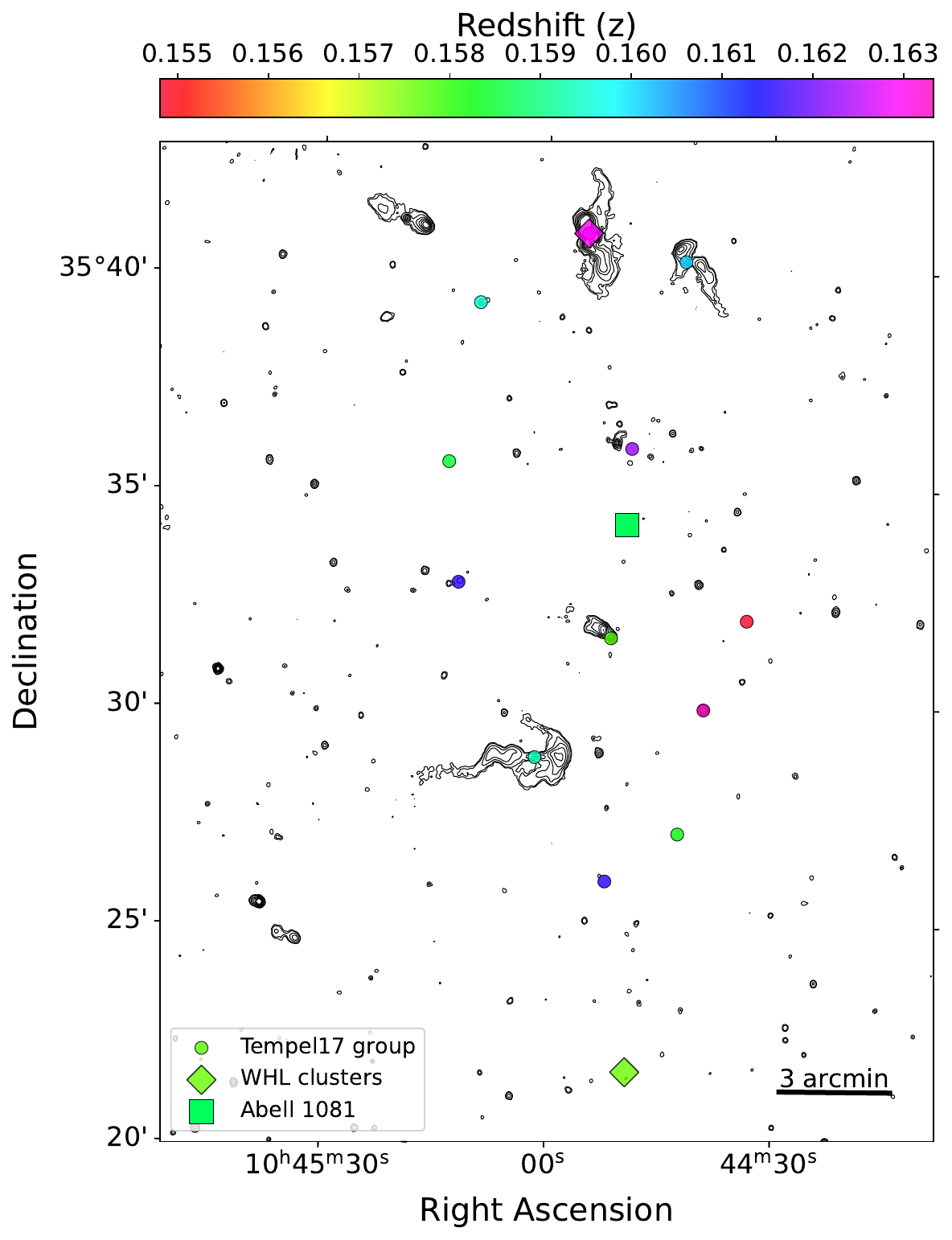}
\caption{Sky distribution of galaxies and clusters in the vicinity of the radio source BAARG overlaid with LoTSS 144~MHz (6\arcsec\ resolution) radio contours. Circular markers represent spectroscopic members of the \citet{Tempel2017} galaxy group (GG\_ID\_325). Diamond symbols indicate the two WHL clusters \citep{whl12}, and the square marks the position of Abell~1081. Marker colours follow a continuous colour scale corresponding to redshift, as shown in the colour bar.}
\label{fig:env}
\end{figure}

The large-scale environment of BAARG is complex and multi-layered, as illustrated in Fig.~\ref{fig:env}. The host galaxy of BAARG lies at $z_{\rm spec} = 0.15939$. The rich galaxy cluster Abell~1081 ($z_{\rm spec} = 0.1588$) is located at a projected angular separation of 5.9\arcmin~ from the BAARG host galaxy. Abell~1081 contains at least 83 member galaxies \citep{Abell1989}, indicating a substantial cluster-scale gravitational potential in close projected proximity to BAARG. In addition, two WHL clusters identified by \citet{whl12} are situated nearby. The cluster WHL\,J104449.3$+$352139 ($z_{\rm spec} = 0.15759$) lies at a projected separation of 7.6\arcmin~ from BAARG, while WHL\,J104454.9$+$354055 ($z_{\rm spec} = 0.16294$) is located 12.1\arcmin~ away. Thus, three cluster-scale systems are present within $\sim$12\arcmin~ of the BAARG host galaxy, all within a narrow redshift interval.
BAARG itself is a member of a galaxy group (ID$=$325) identified by \citet{Tempel2017} using spectroscopic SDSS data. This group contains 12 spectroscopically confirmed members spanning a redshift range $0.15480 \leq z \leq 0.16332$. The redshift distribution of the group overlaps with those of Abell~1081 and both WHL clusters. All quoted redshifts are spectroscopic measurements from the Sloan Digital Sky Survey (SDSS). The total redshift spread across the group and neighbouring cluster systems is  $\Delta z \approx 0.0057$, corresponding to a line-of-sight velocity interval of  $\sim 1700$--$2000$~km~s$^{-1}$ at $z \simeq 0.16$.  This exceeds the characteristic internal velocity dispersions of galaxy clusters, which are typically of order $\sim 10^{3}$~km~s$^{-1}$ \citep[e.g.][]{Bahcall1988}.  Even extremely massive systems rarely exhibit dispersions much above  $\sim 1400$~km~s$^{-1}$ \citep[e.g.][]{Allen2011}. Velocity widths approaching $\sim 2000$~km~s$^{-1}$ are therefore more naturally associated with redshift-space elongation or superposed cluster-scale structures rather than a single dynamically relaxed halo \citep{Bahcall1988}. 
We thus interpret the observed redshift distribution as indicative of a dynamically complex, multi-halo environment. Taken together, the spectroscopic redshift proximity of the galaxy group and neighbouring cluster-scale systems, combined with the pronounced morphological asymmetry of the radio source, indicates that BAARG is embedded in a non-standard environment characterised by overlapping halo-scale structures rather than an isolated, virialised system. Such dynamically complex, multi-halo environments are expected to host large-scale accretion and virial shocks associated with structure formation.

\section{Discussion}
The morphology of BAARG shows a sharp, bow-like structure on one side, suggestive of compression, while the opposite side develops into an extended, diffuse tail. The bow-like structure extends to a projected length of $\sim$280 kpc on one side, significantly larger than the extent of the western lobe ($\sim$115 kpc). This scale difference suggests that the feature is unlikely to be explained solely by backflow or bridge emission typically seen in FR\,II RGs. The available spectral index images (144 to 1400~MHz) show a tentative trend of spectral index being flat at the extreme western end or bow region, but not at the emission peak within the sector region (see Fig.~\ref{fig:rgbm-specin}). We therefore consider a scenario in which the source is interacting with the surrounding medium, potentially shaped by large-scale environmental effects. In this model, we primarily consider that the movement of the BAARG host-galaxy, along with its circum-galactic medium, is supersonic relative to the ambient medium. The resulting large-scale shock, highlighted by the jet-supplied plasma, is thus larger than the lobe and covers the host galaxy in the shock-shell.

Here it is worthwhile to point out, that an alternate scenario could also be possibly imagined where the source is a wide angle tail (WAT) or a bent-tail radio galaxy.
Complex radio galaxies with bent-tail morphologies can arise through several mechanisms, including jet precession, buoyancy, ram pressure, and irregularities or bulk motions in the ambient medium \citep[e.g.][]{axe-shaped-Sudheesh2025, peculiar-Kumari2024, dancing-ghosts-Velovic2023, complex-sources-LoTSS-Horton2025, OdeaWAT2023, TGSS-WAT-Bhukta2022}. 
In the case of BAARG, the ram pressure from the environment could be pushing the bent tails (south-eastern tail of the eastern jet/plume, southern part of the bow and the northern part of the bow) away from the observer and to the east. The asymmetry and turbulence of its complex large-scale environment could explain different orientations of all the tails.

\vspace{-0.5cm}
\subsection{Shock formation during galaxy infall and its observational manifestation}      
In hierarchical structure-formation models, massive galaxy clusters grow by accreting smaller sub-clusters, galaxy groups, and individual galaxies. During the infall of an individual galaxy into a cluster, particularly along a filament, the infall velocity can increase substantially, leading to the formation of a shock ahead of the galaxy. Although some observational evidence has been found in X-ray observations \citep[e.g.][]{shock-X-ray-Irwin1996}, 
the same is unknown in radio bands. Such shocks are clearly visible in simulation studies through electron-density and temperature maps \citep[e.g.][]{shock-simulation-Stevens1999}. For example, simulations of NGC~1404 falling into the Fornax cluster show a shock forming during the first pericentric passage \citep[see electron density map in Fig.~7 and temperature map in Fig.~8 of ][]{cluster-galaxy-shock-simulation-Sheardown2018}. It remains unclear why such shocks have not yet been directly detected in radio observations. However, in the case of BAARG, the presence of an active radio jet at the right epoch (first pericentric passage) and with a favourable orientation may have helped the detection of such a bow-shaped structure around the radio lobe. The small-scale radio jets in BAARG resemble FR\,I-like plumes and show comparable brightness on both sides of the host galaxy, suggesting weak relativistic beaming and a large inclination angle. This is consistent with a near plane-of-sky orientation; a quantitative estimate based on jet-sidedness is presented in Appendix.~\ref{sec:jet_incl}.

Furthermore, the motion of the host galaxy is also likely to be predominantly in the plane of the sky, allowing the jet or plume to interact with the shock near its apex. Had the jet been significantly misaligned, BAARG would instead appear as an RG with a bent-lobe on the side facing the incoming ambient medium. If either the jet or motion of the galaxy were oriented close to the line of sight, the system would likely be observed as a head-tailed or core-halo-type galaxy, with shock-related emission blended with lobe emission. If the jet/lobe is faint, then the hemispherical shock-shell may also appear like an edge-brightened ring (see sec.~\ref{sec:implications}). In the aforementioned simulation study, the shock becomes indistinguishable roughly one billion years after the first pericentric passage. Given that radio-jet activity itself is a relatively short-lived phenomenon (lasting only a few tens of millions of years), BAARG likely represents a rare temporal coincidence for its detection in the radio band. 

\vspace{-0.5cm}
\subsection{Mach number and infall velocity estimates} 
As an order-of-magnitude estimate under a simplified scenario, we derive the approximate infall velocity of the host galaxy through the cluster medium. Assuming that the appearance of the typical arc-shaped morphology ahead of the host galaxy is due to bow shock (as argued before), one can approximately estimate the Mach number from the Mach angle (i.e., the half–angle of the Mach cone, between the flow direction and the Mach wave surface) using the expression $ M = 1/{\sin \theta}$, where $\theta$ is the Mach angle. From Fig.~\ref{fig:Bow-and-Arrow-LoTSS-on-BASS}a, the Mach angle ($\theta$) could be measured approximately as $\theta \approx 25^{\circ}$, which yields a Mach number $M=2.366$. Using the expression for adiabatic sound speed  $c_s =\sqrt{\frac{\gamma k_B T}{\mu m_p}}$, where $k_B$ is the Boltzmann constant, $m_p$ is the mass of a proton, $\gamma$ is the adiabatic index (for monatomic gas = $\frac{5}{3}$), $\mu$ is the mean molecular weight (for fully ionized ICM with cosmic abundances, $\mu \sim 0.60$), and $T$ is the temperature of ICM. For a typical ICM temperature in the range $\sim (10^7 - 10^8) K$, this gives the infall speed of the host galaxy onto the cluster medium to be in the range $\sim$ (1130 - 3580) ~km~s$^{-1}$, which is consistent with the typical supersonic infall speed of the galaxy onto the cluster \citep[e.g.][] {infall-shock-Machacek2005}. These estimates are consistent with a scenario in which the bow-shaped structure arises from a shock associated with the motion of the host galaxy through the cluster medium.


\vspace{-0.5cm} 
\subsection{Implications for radio galaxy morphology and future searches}\label{sec:implications}
Taking BAARG as a motivating example, we outline several possible directions for future investigation. First, given its partial similarity to HyMoRSs \citep{HyMoRS-Gopal2000}, deeper observations of well-selected HyMoRS samples may reveal analogous BAARG-like features. Second, the bow structure in BAARG may be detectable because both the jet and the motion of the host galaxy could be close to the plane of the sky. In cases where the motion is aligned with the line of sight and the jet/lobes are not strong, the bow-structure may appear similar to ORCs \citep{2022Norris, RAD-ORC-Hota2025}. This suggests that the motion of ORC host galaxies may be worth investigating further. Third, the apparent coincidence of galaxies (G1 \& G2) with the emission peaks in the S-shaped structure suggests a potential jet–galaxy interaction scenario \citep[JGI;][references therein]{Hota2022RAD12}; however, spectroscopic redshift measurements are required to establish any physical association.
More generally, deep low-frequency surveys such as LoTSS and future facilities, including the Square Kilometre Array Observatory (SKAO), are likely to reveal additional sources with similar large-scale, low-surface-brightness features. BAARG highlights the potential of such datasets to probe interactions between radio galaxies and their environments, and may serve as an alert for astronomers and citizen scientists alike to discover new cases for follow-up investigations.

\vspace{-0.6cm}
\section*{Acknowledgements}
\begin{small}
The authors are grateful to the anonymous reviewer for carefully reading the manuscript and providing insightful comments, particularly, regarding the aspect of wide angle tail (WAT) scenario. AH acknowledges the University Grants Commission (UGC, Ministry of Education, Govt. of India) for his monthly salary since June 2014. SG acknowledges the Inter-University Centre for Astronomy and Astrophysics (IUCAA) for associateship during which this manuscript was drafted. 
LOFAR data products were provided by the LOFAR Surveys Key Science Project (LSKSP; \url{https://lofar-surveys.org/}) and were derived from observations with the International LOFAR Telescope (ILT). LOFAR \citep{vanHaarlem2013} is the Low Frequency Array designed and constructed by ASTRON. It has observing, data processing, and data storage facilities in several countries, which are owned by various parties (each with its own funding sources), and which are collectively operated by the ILT foundation under a joint scientific policy. The efforts of the LSKSP have benefited from funding from the European Research Council, NOVA, NWO, CNRS-INSU, the SURF Co-operative, the UK Science and Technology Funding Council and the J\"{u}lich Supercomputing Centre. We acknowledge the use of the DESI Legacy Imaging Surveys (\url{https://www.legacysurvey.org/acknowledgment/}). This paper is based [in part] on data from the Hyper Suprime-Cam Legacy Archive (HSCLA; \url{https://hscla.mtk.nao.ac.jp/doc/acknowledgements/}). 

\end{small}
\vspace{-0.5cm}
 \section*{Data Availability}
 \begin{small}
This research has made use of all publicly available astronomy data. LoTSS and BASS data can be obtained via \url{https://lofar-surveys.org/releases.html} and \url{https://www.legacysurvey.org/bass/}.  
\end{small}

\vspace{-0.5cm}
\bibliographystyle{mnras}
\bibliography{ref}

\appendix

\section{Estimates of jet orientation}\label{sec:jet_incl}
To estimate the inclination of the jet axis with respect to the line of sight, we adopt the jet-sidedness formalism of \citet{Hocuk.Barthel}, widely used in similar studies \citep[e.g.][]{Sethi2024}. Assuming intrinsically symmetric jets, the inclination angle ($\theta$) is constrained using the observed jet-to-counterjet flux density ratio ($J$) and the jet bulk velocity ($\beta_j$). Observational and modelling studies indicate that kpc-scale jet velocities typically lie in the range $\sim$0.5c--0.7c \citep[e.g.,][]{Wardle.Aaaron.jet.speed, Mullin2009}. To account for uncertainties, we consider a range of plausible jet velocities.

The inclination angle $\theta$ is given by:
\begin{equation}\label{eq:jet-angle}
\theta = \arccos \left[ \frac{1}{\beta_j} \, \frac{s - 1}{s + 1} \right],
\end{equation}
where $s = J^{1/(2 - \alpha)}$ and $\alpha$ is the spectral index of the jet emission. From the 144~MHz LoTSS~DR2 image, we estimate a jet-to-counterjet ratio of $J = 1.8$, following \citet{Mullin2009}, and adopt an average jet spectral index of $\alpha = -0.6$, as indicated by the spectral index maps (Fig.~\ref{fig:rgbm-specin}).

Using these values, we derive the $\theta$–$\beta_j$ relation (cf. \citealt{Sethi2024}). The corresponding inclination angles are $\theta = 83^\circ$, $79^\circ$, and $67^\circ$ for $\beta_j = 0.99c$, $0.6c$, and $0.3c$, respectively. Across this range of plausible jet velocities, the inferred inclination remains large, indicating that the kpc-scale jets of BAARG are likely oriented at a substantial angle to the line of sight, consistent with a near plane-of-sky configuration.

\section{Images}\label{sec:app-ims}

\begin{figure*}
\centering
\includegraphics[scale=0.29]{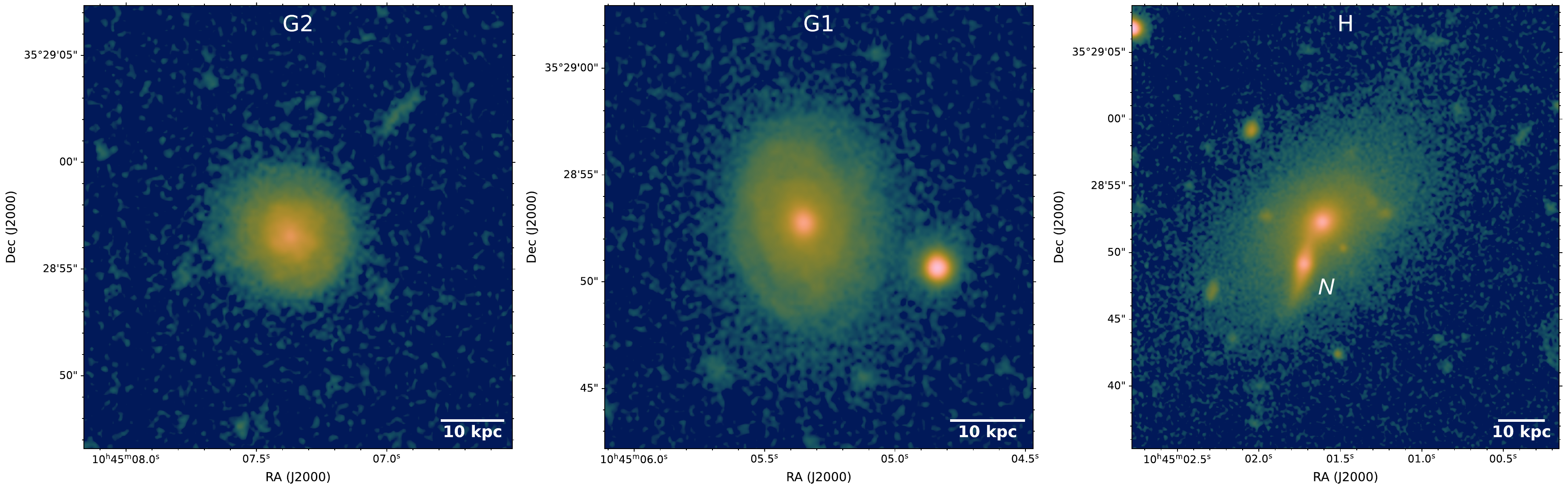}
\caption{ HSCLA optical (G band) images of galaxies G2, G1, and the BAARG host galaxy (H) as marked in Fig.~\ref{fig:Bow-and-Arrow-LoTSS-on-BASS}, along with its neighbouring (in projection) galaxy (N). For more details, see Sec.~\ref{sec:Bow}.}
\label{fig:O_HSCLA}
\end{figure*}

\begin{figure*}
\centering
\includegraphics[scale=0.38]{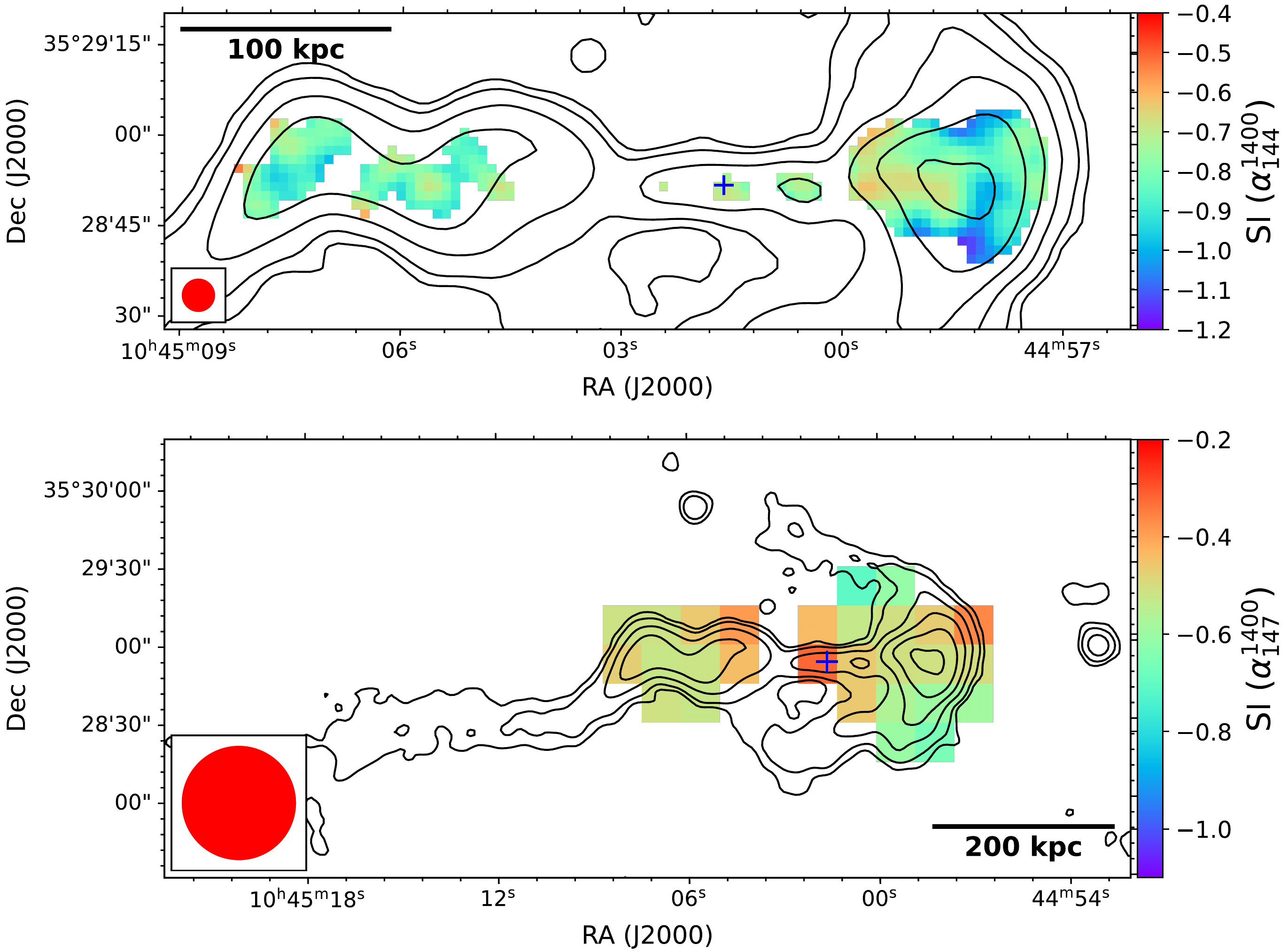}
\caption{We present high- and low-resolution spectral index maps spanning $\sim$\,150–1400~MHz. Due to sensitivity limitations, the full extent of BAARG is not recovered; these maps are therefore indicative and reflect the structures detected in FIRST, NVSS, and TGSS. In both panels, black contours correspond to the LoTSS 6\arcsec\ image at the same levels as in Fig.~\ref{fig:Bow-and-Arrow-LoTSS-on-BASS}. The host galaxy is marked with a blue `+' symbol, and the beam is shown in red in the lower-left corner.
\textit{Upper:} High-resolution spectral index map derived from FIRST and LoTSS 6\arcsec\ images, using emission above $3\sigma$. 
\textit{Lower:} Low-resolution spectral index map from the SPIDX database \citep{SPIDX-deGasperin2018}, based on NVSS and TGSS data.}
\label{fig:rgbm-specin}
\end{figure*}

\bsp	
\label{lastpage}
\end{document}